\documentstyle[twoside,fleqn,epsf,espcrc2]{article}

\newcommand{\AmS}{{\protect\the\textfont2
  A\kern-.1667em\lower.5ex\hbox{M}\kern-.125emS}}

\newcommand{\gsim}{{\protect
  \kern.18em\lower.5ex\hbox{$\stackrel{>}{\sim}$}\kern.25em}}
\newcommand{\lsim}{{\protect
  \kern.17em\lower.5ex\hbox{$\stackrel{<}{\sim}$}\kern.23em}}

\newcommand\be{\begin{equation}}
\newcommand\ee{\end{equation}}
\newcommand\bea{\begin{eqnarray}}
\newcommand\eea{\end{eqnarray}}

\title{
Some aspects of simulation algorithms for dynamical fermions 
        \thanks{
                Plenary talk given by K.J. at the workshop 
                ``Accelerating Fermion Algorithms'', J\"ulich,
                 February 21st to 23rd, 1996.}}

\author{Karl Jansen$^a$, 
        Beat Jegerlehner$^b$ 
        and Chuan Liu$^a$\\ \vspace{0.5cm} 
        $^a$Deutsches Elektronen-Synchroton DESY, \\
        $^{ }$Notkestr. 85, 22603 Hamburg, Germany\\ \vspace{0.5cm}
        $^b$Max-Planck-Institut f\"ur Physik, \\
        $^{ }$F\"ohringer Ring 6, 80805 M\"unchen, Germany}

\begin{document}

\begin{abstract}
Three topics concerning fermion simulation
algorithms are discussed: 1.) A performance comparison of 
the multiboson technique to simulate dynamical fermions 
and the Kramers equation algorithm, 2.)
the question of reversibility in the Hybrid Monte Carlo algorithm
and 3.) the implementation of Symanzik's improvement program for
dynamical Wilson fermions.
\end{abstract}

\maketitle

\section{Introduction}

Wilson's suggestion to formulate QCD on an euclidean space-time lattice
opened the path to non-perturbative and first principle investigations
of strong interactions. It was soon realized that for pure Yang-Mills
theory numerical simulations are a powerful tool to address several 
questions in this theory. In particular, numerical simulations can
build a bridge between
the high energy regime, where perturbation theory is valid, to scales
where non-perturbative effects set in. 

However, 
as the title of this workshop already suggests,
simulations with dynamical
fermions have still to be accelerated in order to obtain results with
small enough error bars in a reasonable amount of time. 
Therefore, 
in the last years we have seen a revived interest in a better understanding 
of the existing algorithms (the old work horses) as we have seen 
 new developments.

Such a development is the multiboson technique 
to simulate dynamical fermions \cite{martin} and we will
in section 2 compare its performance to the Kramers equation algorithm 
\cite{horowitz,kramers,kramersboson}.
Section 3 is concerned with the question of lack of reversibility in the Hybrid
Monte Carlo algorithm \cite{kramers} which may cause potential problems 
with the detailed balance condition in practical simulations.
The 4th section 
will contain a short account of the implementation of 
Symanzik's improvement program for Wilson fermions. The overhead due to
the addition of the extra Sheikholeslami-Wohlert (SW) term 
\cite{clover} in the lattice
action will
be discussed.
Note that in sections 2 and 3 the gauge group is chosen to be SU(2) while
in section 4 it is SU(3).

Space limitations do not allow to give here a full account of all the
notations and background. For this we have to refer to the original
literature \cite{kramers,kramersboson,beat,hmcsw}.

\section{Performance comparison of the 
multiboson technique to simulate dynamical fermions and the 
Kramers equation algorithm}

In this section, the performance of two 
algorithms is compared. 
The first one 
goes back
to a suggestion of Horowitz \cite{horowitz} and is called the Kramers
equation algorithm. 
In a free field analysis it gives the same value of the dynamical
critical exponent, $z=1$, as the Hybrid Monte Carlo (HMC) algorithm
\cite{tony}. However, whereas in the HMC algorithm this value of $z$ is
only reached in the limit of a large number of molecular dynamics 
steps, the Kramers equation algorithm needs only one step. 
The second algorithm is based on a new technique that makes use of
the fact that the QCD partition function can be rewritten as a path
integral of --in principle-- infinitely many boson fields with interactions 
given by the fermion matrix on the lattice. 
Fortunately enough, it turned out that
for practical simulations only a moderate number of these boson fields
is required to obtain a good approximation of the QCD partition function
\cite{bunk}.
The tests of the Kramers equation algorithm and the multiboson technique
have been performed for Wilson fermions.                            
For a detailed description and investigation of the two
algorithms see \cite{beat} and \cite{kramers}. 
For alternative realizations of the multiboson technique see \cite{multiboese}
and for a review \cite{forcrand}.
Let us here only emphasize
the improvements over the ``bare'' algorithm that we have installed. 
Starting with the Kramers equation algorithm, we 
\begin{itemize}
\item installed standard even-odd preconditioning \cite{precond}
\item and used a particular integration scheme that allows for
      larger step sizes \cite{sexy}.
\end{itemize}

For the multiboson technique we
\begin{itemize}
\item used again even-odd preconditioning,  
\item introduced a normalization factor for the fermion matrix that lifts
the lowest eigenvalue while keeping the largest one below one 
\cite{multiboese,beat} 
\item and made a careful investigation of the optimal choice of the
mixing ratio of heatbath and over-relaxation updates \cite{beat}.
\end{itemize} 

Let us mention that each of the above improvements 
 may lead to substantial factors 
of accelerating the program. 
We also looked at 
the chronological
extrapolation method to find better starting vectors for the conjugate
gradient algorithm \cite{brower}.
However, due to the use of large step sizes in the
Sexton-Weingarten integration scheme, we did not find a further acceleration
of the program in our case. 

The algorithm and coupling parameters used for the tests may be found in 
\cite{kramersboson}. 
We worked at a rather large pion to $\rho$-mass ratio 
of $0.95$. 
All numerical simulations have been performed on the Alenia Quadrics
(APE) massively parallel machines. In table~1, we give the time 
$\tau$ in real seconds that 
the algorithms need to generate an independent configuration.
The subscript $b$ stands for the multiboson technique and $k$ for
the Kramers equation algorithm. 
The computationally most expensive part
in the Kramers equation algorithm is the conjugate
gradient method for the matrix inversion. This inversion, on the other
hand, is dominated by fermion matrix $Q$ \cite{kramers} 
times vector $\phi$ operations,
denoted by $Q\Phi$.
Also for the bosonic algorithm similar floating point operations
dominate the program \cite{bunk,beat}.
For this reason we give the autocorrelation time
$\tau[Q\Phi]$
in units of $Q\Phi$ in columns 3 and 4 which is a more machine 
independent measure of the performance. 

\begin{table*}[hbt]
\caption{Performance comparison
of the Kramers equation and the boson algorithms. Subscripts $b$ and $k$
stand for the multiboson technique and the Kramers equation algorithm,
respectively.
}
\vspace{2mm}
\label{tab:table3}
\begin{tabular*}{\textwidth}{@{}l@{\extracolsep{\fill}}ccccc}
\hline
                   \multicolumn{1}{c}{Lattice}
                 & \multicolumn{1}{c}{machine size}
                 & \multicolumn{1}{c}{$\tau_b [Q\Phi]$}
                 & \multicolumn{1}{c} {$\tau_k [Q\Phi]$}
                 & \multicolumn{1}{c} {$\tau_b [sec]$}
                 & \multicolumn{1}{c} {$\tau_k [sec]$}
\\
\cline{1-6} \cline{1-6}
\hline \hline
$6^312$ & Q1 [8 nodes]    & $ 14000(800)$ & $ 21000(4000)$
                          & $ 298(20)$  & $ 480(100) $ \\
$8^312$ & Q1 [8 nodes]    & $ 36000(2250)$ & $ 17000(5000)$
                          & $ 1781(112)$ & $ 979(300) $ \\
$16^4$ &  QH2 [256 nodes] & $ 56000(18600)$ & $ 26000(11000)$
                          & $  990(330)$ & $540(230)$ \\
\hline
\vspace{-5mm}
\end{tabular*}
\end{table*}

We see that within the error bars for the autocorrelation time,
both algorithms show a comparable performance. 

\section{Reversibility}

In order for the HMC algorithm to
fulfill the detailed balance condition, 
reversibility
of the equations of motion has to hold. 
However, problems
with the reversibility condition have been encountered and in this section 
some aspects of these problems will be discussed.

In the HMC algorithm the fields are evolved according to a 
non-linear first order
stochastic differential equation. In the continuum, such differential
equations are candidates for describing the time evolution of 
a classical chaotic system. 
Of course, for the computer, the differential equation has to be
discretized. Nevertheless, one might expect that for small enough 
discrete step sizes one is close enough to the continuum and that one 
would observe a chaotic behaviour.
To test this proposal, let us
define a quantity, $||dU||$,  to measure the
difference between two gauge field configurations
\be \label{dudef}
\| dU\|^2 = {1 \over 4\Omega}
\sum_{x,\mu,a} (U_\mu^a (x) - V_\mu^a (x))^2\; .
\ee
Here $U^a_{\mu}(x),V^a_{\mu}(x)$ are two SU(2)
gauge link variables with lattice point
index $x$, direction $\mu$ and group index $a$.
$\Omega$ is the lattice volume.

To show that the suggested chaotic behavior is really a property of
Hamilton's equations of motion, we proceeded
in the following way. Given a
gauge field configuration obtained in the course of some run, we
added a small noise $\delta U_{\mu}(x)$ to the
gauge field variable $U_{\mu}(x)$, such that
$V_{\mu}(x)=U_{\mu}(x)+\delta U_{\mu}(x)$.
Then, we took both configurations and iterated
them according to the leapfrog integration scheme used in the HMC algorithm.
We measured $\|dU\|$ after some number of steps $N_{md}$
in the leapfrog integration. If the system is chaotic,
we expect that asymptotically ($\epsilon N_{md} \gg 1$)
\be \label{revexp}
\|dU\| =A e^{\nu \epsilon N_{md}}.
\ee
In eq.~(\ref{revexp}) $\nu$ is the --to be determined-- Liapunov exponent,
characterizing a chaotic system and $\epsilon$ is the step size used
in the program.

Indeed, in \cite{kramers} the asymptotic exponential grow of
$\|dU\|$ was verified and a positive Liapunov exponent $\nu \approx 0.75$
was extracted. 
To see whether this effect also happens in a real Monte Carlo
run with dynamical Wilson fermions using the HMC algorithm we did 
the following:
When a trajectory has been integrated, 
we reversed the time and integrated back,
measuring $\| dU\|$ using the initial configuration
before starting the leapfrog integration and the final configuration at the
end of the reversed trajectory.
By fixing the step size $\epsilon =0.03$, we obtained $\| dU\|$ as a function
of the trajectory length $\epsilon N_{md}$. Indeed, we find
again a linear
behavior of $lg(\|dU\|)$ with a Liapunov exponent around $0.75$.     
A similar observation has been made in the case of SU(3) gauge fields
\cite{wuppertal}.

An important consequence of this observation is that the necessary condition
of reversibility in the detailed balance proof of the HMC algorithm may 
be violated in practical simulations. Rounding errors appearing in
the course of the simulation are blown up exponentially and
destroy the reversibility of Hamilton's equations of motion. 
We tried to estimate, how serious this effect is for a real simulation. 
We ran two HMC programs simultaneously: One with 32 bit arithmetic but
with all scalar products, dot products etc. in double precision (which 
mirrors
the case of the APE version of our program). The second one running with
complete 64 bit arithmetic. 
The Metropolis decision in the HMC algorithm is on $\Delta H$, i.e. the
difference of the start and the end Hamiltonian. 
We measured the difference $\delta(\Delta H)$ between $\Delta H$ of the
program with 32 bit and 64 bit precision. 
This was done for each step in the evolution 
$\delta(\Delta H)_{step}$ and for a complete trajectory, 
$\delta(\Delta H)_{traj}$. We find that on a $16^4$ lattice 
$\delta(\Delta H)_{traj}$ to be larger than 
$\delta(\Delta H)_{step}$, indicating that indeed 
a growth of the errors due to the chaotic nature is showing up.

Repeating this test on lattices of sizes $8^4$, $12^4$ and $16^4$, we
found that by a simple extrapolation in the lattice size,
for a $32^4$ lattice the ratio
$\delta(\Delta H)_{traj}/\Delta H \approx 0.05$, whereas 
$\delta(\Delta H)_{step}/\Delta H \approx 0.01$.
We believe that a
difference of $\Delta H$ between the 32 bit and the 64 bit
program versions reaching this level, may
lead to observable effects in physical
quantities. 

\section{Implementation of Symanzik's improvement program} 

Following Symanzik \cite{symanzik}, 
in order to cancel the $O(a)$ effects in physical on-shell observables,
it is sufficient for Wilson fermions to add a term suggested by
Sheikholeslami and Wohlert \cite{clover}, henceforth called a SW-term. 
In \cite{letter}, it has been demonstrated 
in the quenched approximation that the $O(a)$ effects are
substantial for Wilson fermions. There it was also shown that for a 
complete cancellation of $O(a)$ effects a non-perturbative determination
of the coefficients multiplying the improvement terms is necessary.
Given the experience of this work, it is a very natural next step 
to also implement the SW-term for dynamical fermions.
The force according to the SW-term can be
derived straightforwardly and it can be shown that even-odd 
preconditioning can be maintained. 
For a detailed description of our implementation and the design
of possible tests of the code see \cite{hmcsw}.

When comparing with the conventional Wilson QCD simulation,
the timing of the version with
the improved fermion action is the crucial piece of information.

\begin{figure}[thp]
\vspace{-9mm}
\centerline{ \epsfysize=8.0cm
             \epsfxsize=8.0cm
             \epsfbox{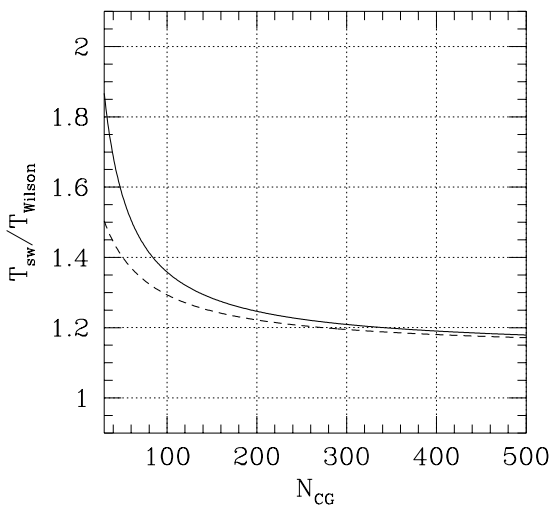}}
\vspace{-15mm}
\begin{center}
\parbox{7.5cm}{\caption{ \label{fig:time}
 The CPU time for the algorithm of the improved action
 is plotted as a function of the number of conjugate
 gradient iterations per step ($N_{CG}$) relative to
 the conventional Wilson fermion simulation.
 The solid and the dashed lines correspond to the choice
 of $t^{(0)}_w=0$ and $t^{(0)}_w=50$ in eq.~\protect{(\ref{eq:time})}
 respectively.
}}
\vspace{-9mm}
\end{center}
\end{figure}
In a typical QCD simulation using molecular dynamics
algorithms,
 the conjugate gradient (CG) iterations dominate
the CPU time of the program.
Therefore, in the limit of number of conjugate
gradient iterations $N_{CG}$ going to infinity,
 the time of the matrix-vector
multiplication is the most crucial part.
In our version of Symanzik improvement,
this multiplication turns out to be
only slightly slower than the conventional
Wilson case by about 15 percent.
The force evaluation, in this case, adds some
overhead to the program which does not
depend on $N_{CG}$. It is
convenient to express all times in units of the
matrix-vector multiplication of the
conventional Wilson case.

To be specific, we propose the
following formula for the time of the program:
\vspace{-0.3mm}
\begin{eqnarray}
\label{eq:time}
T_{Wilson} &=& t^{(0)}_w+(t^{(1)}_w+t^{(2)}_w)N_{CG} \;\;,
\nonumber \\
T_{SW} &=& t^{(0)}_{sw}+(t^{(1)}_{sw}+t^{(2)}_{sw})N_{CG} \;\;.
\vspace{-1mm}
\end{eqnarray}
In the above formula, the coefficient $t^{(2)}_w=2$ is, by definition, the
number of matrix-vector
multiplications needed for each CG iteration in
the conventional Wilson case. The coefficient $t^{(1)}_w$ represents
the cost of other operations for each CG iteration (linear combinations,
inner products etc.). Typically, this coefficient is quite
small. The coefficient $t^{(0)}_w$ is  the
overhead that does not depend on  $N_{CG}$. The value of
$t^{(0)}_w$ depends on the implementation of the program.
However, in the asymptotic region where $N_{CG}$ is large,
the effect of $t^{(0)}_w$ becomes irrelevant.

The quantities $t^{(0)}_{sw}$, $t^{(1)}_{sw}$ and $t^{(2)}_{sw}$
are the corresponding coefficients for the program with
the Sheikholeslami-Wohlert action.
In this case, $t^{(2)}_{sw}=2.3$ is slightly larger than
the corresponding value in the Wilson case. The value of
$t^{(1)}_{sw}=t^{(1)}_{w}$ remains the same and
we find, for our implementation of the SW-term on the APE computer,
the coefficient
$t^{(0)}_{sw}\sim (50+t^{(0)}_w)$ to be larger than that of
the Wilson case.
In Fig.~\ref{fig:time} we plot the CPU time of our program
with improved fermions relative to the
conventional Wilson case as a function
of the  number of CG-iterations per molecular dynamics step.
The solid and dashed curves in the figure correspond to
the choice of $t^{(0)}_w=0$ and $t^{(0)}_w=50$ respectively, which
we consider to be two rather extreme cases.
For typical simulations, the curve should lie between these two.
In any case, we see that, for the most
interesting physical situation in which
$N_{CG} \sim 200$ or more, the simulation with improved fermions
is only about 20\%  slower, quite independent
of the value of $t^{(0)}_w$.

\section{Conclusions}

We presented a performance test of the multiboson technique 
to simulate dynamical fermions and  
the Kramers equation algorithm. We find that these two quite different
simulation techniques for fermions give a comparable performance.
Approaches to make the multiboson technique exact by adding a 
Metropolis step are promising: The number of boson fields can 
be chosen to be small while keeping large acceptance rates
\cite{multiboese}. A detailed performance 
comparison of a version of an exact multiboson program 
and the Kramers equation or HMC algorithm is,
however, not yet available. 

Possible problems with the lack of reversibility in the discretized 
Hamilton's equations of motion have been discussed in section 3.
A rough estimate leads to the conclusion that on lattices with size
$32^4$ on machines with 32 bit arithmetic the lack of reversibility 
can give serious problems. For the Kramers equation algorithm this
problem is substantially reduced, since one only would need one
molecular dynamics 
step. 

Coming back to the title of the workshop, did we now accelerate the
fermion algorithms? The answer is yes and comes from a quite unexpected
direction. The SW-term as required by
Symanzik's improvement program can be implemented straightforwardly 
for Wilson fermions. Fig.~1 demonstrates that 
the slow down of the program is only about 
20\%. On the other hand, by determining the coefficients
multiplying the improvement terms 
non-perturbatively, all $O(a)$ effects in physical on-shell observables can be
eliminated which results in a much faster approach to the continuum limit,
in which we are most interested in.  

\vspace{-3mm}
\section*{Acknowledgement}
It is a pleasure to thank A. Galli, M. L\"uscher, P. Weisz and 
U. Wolff 
for many useful and stimulating 
discussions.

\end{document}